\documentclass[12pt]{article}
\usepackage{amssymb,amsmath} 
\usepackage{graphicx}
\graphicspath{{graphics/}}
\usepackage{latexsym}
\usepackage{subfig}  
\usepackage{booktabs,setspace,caption}  
\usepackage{braket}
\usepackage{wasysym,verbatim}
\usepackage[numbers,sort&compress]{natbib}
\usepackage{textcomp}
\usepackage[utf8]{inputenc}
\usepackage{mathrsfs}
\usepackage[inner=2.9cm,outer=2cm]{geometry}
\usepackage[toc]{appendix}
\usepackage{color,soul} 
\usepackage{datetime}
\usepackage[
      colorlinks=true,
      linkcolor=darkblue,  
      urlcolor=black,    
      filecolor=blue,     
      citecolor=darkgreen, 
      linktocpage=true,
      pdfstartview=FitV,
      bookmarksopen=true    
      ]{hyperref}

\definecolor{darkblue}{rgb}{0.2, 0, 0.8}
\definecolor{darkgreen}{rgb}{0.2, 0.71, 0}

\numberwithin{equation}{section}

\newenvironment{changemargin}[2]{%
\begin{list}{}{%
\setlength{\topsep}{0pt}%
\setlength{\leftmargin}{#1}%
\setlength{\rightmargin}{#2}%
\setlength{\listparindent}{\parindent}%
\setlength{\itemindent}{\parindent}%
\setlength{\parsep}{\parskip}%
}%
\item[]}{\end{list}}

\begin{document}  
%%%%%%%%%%%%%%%%%%%%%%%%%%%%%%%%%%%%%%%

%%%%%%% title page %%%%%%%%%

\begin{titlepage}

\begin{flushright}
{\tt \small{IFT-UAM/CSIC-17-080}} \\
%{\tt \small{IPhT-T16/070}} \\
\end{flushright}

\vspace*{1.2cm}

\begin{center}
{\Large {\bf{One Thousand and One Bubbles} }} \\

\vspace*{1.2cm}
\renewcommand{\thefootnote}{\alph{footnote}}
{\sl\large Jes\'us \'Avila$^{\text{\quarternote}}$, Pedro F.~Ram\'{\i}rez$^{\text{\eighthnote}}$ and Alejandro Ruip\'erez$^{\text{\eighthnote}}$ }\footnotetext{{j.avila.c[at]csic.es  ; p.f.ramirez[at]csic.es ; alejandro.ruiperez[at]uam.es }}
\bigskip

$^{\text{\quarternote}}$Instituto de Ciencia de Materiales de Madrid ICMM/CSIC\\
C/ Sor Juana In\'es de la Cruz, 3, C.U. Cantoblanco, 28049 Madrid, Spain\\

\bigskip

$^{\text{\eighthnote}}$Instituto de F\'isica Te\'orica UAM/CSIC \\
C/ Nicol\'as Cabrera, 13-15, C.U. Cantoblanco, 28049 Madrid, Spain\\

\setcounter{footnote}{0}
\renewcommand{\thefootnote}{\arabic{footnote}}

\bigskip

\bigskip

\end{center}

\vspace*{0.1cm}

%\noindent\rule{16cm}{0.4pt}
\begin{abstract}  
\begin{changemargin}{-0.95cm}{-0.95cm}
%\noindent  
We propose a novel strategy that permits the construction of completely general five-dimensional microstate geometries on a Gibbons-Hawking space. Our scheme is based on two steps. First, we rewrite the bubble equations as a system of linear equations that can be easily solved. Second, we conjecture that the presence or absence of closed timelike curves in the solution can be detected through the evaluation of an algebraic relation. The construction we propose is systematic and covers the whole space of parameters, so it can be applied to find all five-dimensional BPS microstate geometries on a Gibbons-Hawking base. As a first result of this approach, we find that the spectrum of scaling solutions becomes much larger when non-Abelian fields are present. We use our method to describe several smooth horizonless multicenter solutions with the asymptotic charges of three-charge (Abelian and non-Abelian) black holes. In particular, we describe solutions with the centers lying on lines and circles that can be specified with exact precision. We show the power of our method by explicitly constructing a 50-center solution. Moreover, we use it to find the first smooth five-dimensional microstate geometries with arbitrarily small angular momentum. 
\end{changemargin}
\end{abstract} 
%\noindent\rule{16cm}{0.4pt}

\end{titlepage}

%%%%%%%%%%%%%%%%%%%%%%%%%%%%%%%%%%%%%
\setcounter{tocdepth}{2}
%the line above sets the depth of the table of contents. {2} means it will display section and subsections only. 
{\small
\setlength\parskip{-0.5mm} 
\noindent\rule{15.7cm}{0.4pt}
\tableofcontents
\vspace{0.6cm}
\noindent\rule{15.7cm}{0.4pt}
}

%%%%%%%%%%%%%%%%%%%%%%%%%%%%%%%%%%%%%%%%%%%%%%%%%%%%%%%%%%%%
%%%%%%%%%%%%%%%%%%%%%%%%%%%%%%%%%%%%%%%%%%%%%%%%%%%%%%%%%%%%
\section{Introduction}

Microstate geometries are solitonic solutions of the equations of motion of supergravity theories. Classical results from General Relativity established that this type of non-singular solutions cannot be accommodated in a four-dimensional spacetime\footnote{See for instance the early work \cite{10.2307/1968759}. A more recent article with emphasis in microstate geometries is \cite{Gibbons:2013tqa}.}, at least when non-Abelian matter fields are absent. Actually, asymptotically flat, spherically symmetric non-Abelian solitons have been known to exist in four and more dimensions for decades, \cite{Bartnik:1988am, Strominger:1990et, Harvey:1991jr, Chamseddine:1997nm, Chamseddine:1997mc, Hubscher:2008yz, Bueno:2014mea, Cano:2017sqy}, although multicenter solutions have only been discovered very recently \cite{Ramirez:2016tqc}. Abelian solitons in supergravity, however, are only possible in five dimensions or more. In that case earlier no-go theorems can be circumvented due to the presence of Chern-Simons topological terms in the action; magnetic fluxes threading non-contractible 2-cycles become effective sources of electric charge, mass and angular momentum. It is for this reason that Abelian microstate geometries require the spacetime manifold to have non-trivial topology.

A set of rules to construct Abelian microstate geometries as supersymmetric solutions of five-dimensional supergravity was discovered in \cite{Berglund:2005vb, Bena:2005va}, where these were conjectured to be related to the classical description of black hole microstates within the context of the \emph{fuzzball proposal} \cite{Mathur:2005zp}. These works generalized earlier results \cite{Lunin:2001jy, Lunin:2002iz, Mathur:2003hj, Lunin:2004uu, Giusto:2004id, Giusto:2004ip, Giusto:2004kj} by making use of the solution generating technique of \cite{Gutowski:2004yv, Bena:2004de}. On the other hand, based on the results of the program for the study of non-Abelian black holes in string theory \cite{Bellorin:2007yp, Meessen:2008kb, Bueno:2015wva, Meessen:2015enl, Ortin:2016bnl, Cano:2016rls, Cano:2017qrq, Meessen:2017rwm}, this technique has been extended to include the construction of non-Abelian microstate geometries in \cite{Ramirez:2016tqc}. In this article we introduce a unified framework, so our discussion can be applied to all five-dimensional supersymmetric microstate geometries on a Gibbons-Hawking base\footnote{The recent solutions of \cite{Fernandez-Melgarejo:2017dme} describing supertubes with non-Abelian monodromies are constructed using harmonic functions with codimension-2 singularities and are not included in our framework.}.

The aforementioned solution generating technique, however, has very limited applications. In few words, this is due to the fact that it is not known how to systematically avoid the presence of Dirac-Misner string singularities or general closed timelike curves (CTCs). We refer to this fact generically as the CTCs problem, since Dirac-Misner strings, if present, can only be resolved if the time coordinate is periodic, which introduces CTCs as well. According to the technique to construct solutions, Dirac-Misner strings are absent when the \emph{bubble equations} are solved, see \eqref{bbeq1}, while the geometry is free of general CTCs when the \emph{quartic invariant} function is positive everywhere, \eqref{eq:quartic}. The problem arises because it is not known how the parameters that specify the solution must be chosen in order to satisfy these two types of constraints. For this reason, the solution generating technique of \cite{Berglund:2005vb, Bena:2005va} is rarely directly applied to find explicit solutions, being effective only for simplified configurations with few centers or with very special relations among the specifying parameters, see \cite{Gimon:2007mha, Bena:2007kg} and references therein. More general solutions have been found by making use of more sophisticated tools, like the merging of several few-center solutions, \cite{Bena:2006kb, Bena:2007qc}, or the link between three-charge supertube configurations and five-dimensional microstate geometries on a Gibbons-Hawking base, \cite{Bena:2008wt, Heidmann:2017cxt, Bena:2017fvm}. Superstrata solutions, which belong to a different class of smooth horizonless solutions of six-dimensional supergravity, deserve a special mention, as they might reproduce the degeneracy of microstates of general three-charge black holes, \cite{Bena:2014qxa, Bena:2016ypk}.

In order to understand the situation better, it is convenient to discuss the origin of these pathologies. Timelike supersymmetric solutions of five-dimensional supergravity have a metric of conformastationary form,

\begin{equation}
ds^2 = f^2 \left( dt + \omega \right)^2 - f^{-1} h_{mn} dx^m dx^n \, .
\end{equation}

\noindent
Here $h_{mn} dx^m dx^n$ is a four-dimensional hyperKähler metric (usually a Gibbons-Hawking space \cite{Gibbons:1979zt, Gibbons:1987sp}) known as the \emph{base space}, while $f$ and $\omega$ are respectively a function and a 1-form defined on this base space\footnote{This metric can be interpreted as a fibration of an $\mathbb{R}$-bundle over a four-dimensional space; the hypersurfaces with constant $t$ define a local section and $\gamma_{mn}=- f^{-1} h_{mn} $ is the projection of the spacetime metric orthogonal to the fibres defined by the horizontal connection $\omega$. On the other hand, the metric induced in that hypersurfaces is $\tilde{g}_{mn}=f^2 \omega_m \omega_n - f^{-1}h_{mn}$. As discussed in \cite{Gibbons:2013tqa}, for microstate geometries the coordinate $t$ is a global time function and the sections it defines are Cauchy surfaces.}. The 1-form $\omega$ must transform as a tensor under coordinate transformations on the base space, since otherwise the hypersurfaces defined by constant values of the coordinate $t$ would not be Cauchy surfaces. As $\omega$ is specified by a differential equation, this just means that we need to satisfy the corresponding integrability condition everywhere. This integrability condition becomes a set of algebraic relations known as the bubble equations. From a physical perspective, this phenomena is related to the frame-dragging generated by the interactions between electric and magnetic sources. These sources have Dirac string singularities, and their elimination from any influence in electromagnetic interactions requires imposing charge quantization. In a similar manner, we can think of the bubble equations as the conditions that require the frame-dragging is also invisible to those string singularities\footnote{Alternatively, one could get rid of the string singularities without solving the integrability condition by interpreting $\omega$ as a connection and solving its defining equation on different patches (as it is done, for instance, for the Dirac monopole). However, the consistency of this construction requires the time coordinate $t$ to be compact, as shown by Misner in \cite{doi:10.1063/1.1704019}.}. These constraints relate the \emph{charge} parameters with the sizes of the non-contractible 2-cycles, and are typically interpreted as restrictions for the latter.

On the other side, the above condition is necessary but not sufficient to ensure $t$ is a global time coordinate. Besides, it is necessary that any hypersurface defined by a constant value of $t$ is a Riemannian manifold with timelike normal vector, so that there are no CTCs. For these solutions, this requires the quartic invariant function to be positive. In the context of BPS microstate geometries, this can be rephrased in terms of the signs of \emph{charge and energy densities} at separated locations. The existence of stationary multicenter supersymmetric solutions is typically related to the cancellation of attractive and repulsive (gravitational and electromagnetic) interactions\footnote{Interestingly, one may notice that the action does not contain any terms introducing ``Lorentz-like'' electromagnetic forces. However, metric-based theories of gravity ``know'' about the existence of those interactions through the coupling of the spacetime metric and the electromagnetic energy-momentum tensor. See for example \cite{Infeld:1940zz}.}. If these cancellations cease to take place the configuration is not truly supersymmetric, and this is reflected in the form of CTCs when trying to describe the solution as such. Hence, the problem is to find configurations for which all charge densities are of the same sign. However, the relation between the parameters specifying the solution and the charge densities, which arise from the interactions of magnetic fluxes threading non-contractible cycles, is rather complex. This is the reason why the solution of this problem has remained unclear.

 %%%%%%%%%%%%%%%%%%%%%%%%%%%%%%%%%%%%%%%%%%%%%%%%%%
 
 \subsection{Results and plan of the paper}
 
 In this article we propose a systematic method to solve the CTCs problem that can be used to find all five-dimensional BPS microstate geometries on a Gibbons-Hawking hyperKähler space. It can be summarized as follows:
 
 \begin{enumerate}
 \item The bubble equations are non-linear and hard to solve if the locations of the centers are taken as the unknowns. However, those can be rewritten as a simple system of linear equations by choosing a different set of unknown variables: the magnetic fluxes. The bubble equations become
 \begin{equation}
 \mathcal{M}  X =  B \, , \qquad
 \end{equation}
 for some symmetric matrix $\mathcal M$.
 \item We conjecture that any solution satisfying the bubble equations is free of CTCs if and only if all the eigenvalues of the matrix $\mathcal M$ are positive. 
 \end{enumerate}
 
When trying to build generic microstate geometries, all parameters specifying the solution can be treated in an equal footing. Therefore, there is no reason to consider the charge parameters more fundamental than the size of the bubbles. We begin with a description of the parameter space in section \ref{sec:parspace}. Then, we consider the CTCs problem in section \ref{sec:CTCproblem}. In particular, in section \ref{solvbbeqs} we rewrite the bubble equations as a linear system with the same number of equations than variables, while in section \ref{sec:CTC} we expose our conjecture and discuss evidence in its support. In section \ref{sec:scaling} we discuss the application of our method to the construction of scaling solutions, describing how the introduction of non-Abelian fields strongly enriches the spectrum of this type of solutions. Afterwards, in section \ref{sec:thousand}, we put our method in practice and describe some solutions with properties and characteristics previously absent in the literature. For instance, as a striking result, we are able to find smooth horizonless five-dimensional solutions with arbitrarily small angular momentum. Those had not been discovered so far and they were even thought to be nonexistent. We also describe some solutions with the centers lying on a circle or a line whose parameters can be specified with complete accuracy (to the best of our knowledge, all the explicit microstate geometries with several centers known so far can only be obtained approximately). Last but not least, we find a solution with 50 centers that contains more than a thousand 2-cycles, which inspired the title of this article. We conclude in section \ref{sec:conclusions} making some final comments and proposing future lines of research to exploit the possibilities that our method of construction offers. The appendices contain the details about the theory and the solution generating technique we use.

 %HHHHHHHHHHHHHEEEEREEEEEEEE

%
%Blablabla of microstate geometries, including several references and state of the art. 
%
%There are construction rules and explicit examples of different kinds. One major obstacle in describing multicenter solutions are Dirac-Misner strings. Usually these are algebraic equations constraining the distances between the centers, which complicates the task. There are (n-1) equations for n(n-1)/2 variables which are in principle independent but in reality are not, as they must be real numbers satisfying triangle inequalities. Moreover, these are three-dimensional distances and this means the number of independent distances is only 3(n-2). This is why usually people work with the points aligned (also omega...), in which case the bubble eqs become a system with the same number of equations and unknowns. However the construction of solutions is still hard, as the system is not linear and the solutions are usually complex or negative.

%%%%%%%%%%%%%%%%%%%%%%%%%%%%%%%%%%%%%%%%%%%%%%%%%%%%%%%%%%%%
%%%%%%%%%%%%%%%%%%%%%%%%%%%%%%%%%%%%%%%%%%%%%%%%%%%%%%%%%%%%
%%%%%%%%%%%%%%%%%%%%%%%%%%%%%%%%%%%%%%%%%%%%%%%%%%%%%%%%%%%%
\section{The parameter space and its restrictions}
\label{sec:parspace}

This section is devoted to the description of the space of parameters of supersymmetric five-dimensional microstate geometries and the restrictions that physical configurations must fulfill. 

We work in the theory of five-dimensional $\mathcal{N}=1$ Super-Einstein-Yang-Mills (SEYM) supergravity\footnote{This theory is obtained by gauging a non-Abelian subgroup, typically $SU(2)$, of the group of isometries of the scalar manifold and its real special structure. See \cite{Ortin:2015hya, Meessen:2015enl}.}. The bosonic field content of this theory is given by the metric, three Abelian vector multiplets and a $SU(2)$ triplet of non-Abelian vector multiplets. If the latter are truncated, the STU model of supergravity is recovered, the theory in which five-dimensional Abelian microstate geometries are constructed. Our scheme, therefore, is completely general and accommodates both classes of solutions, Abelian and non-Abelian. The appendices contain a workable description of the theory and how this type of solutions are constructed.

The theory of SEYM supergravity can be embedded in string theory, as recently shown in \cite{Cano:2016rls, Cano:2017sqy, Cano:2017qrq}. It emerges from the toroidal compactification of heterotic supergravity, followed by a convenient truncation that eliminates half of the original supercharges.

Microstate geometries are completely specified by a set of harmonic functions in three-dimensional Euclidean space\footnote{It is most remarkable that the equations of motion of the non-Abelian sector can be linearized when the \emph{multi-colored dyon ansatz} is employed, \cite{Ramirez:2016tqc, Meessen:2017rwm}.} and some restrictions that strongly constrain the parameter space. Let us first introduce the parameter space and discuss the constraints afterwards. We distinguish the harmonic functions in the Abelian sector

\begin{equation}
\label{eq:Abmulti}
H=\sum_{a=1}^n \frac{q_a}{r_a} \, , \qquad
\Phi^i=\sum_{a=1}^n \frac{k^i_a}{r_a} \, , \qquad
L_i=l^i_0+\sum_{a=1}^n \frac{l^i_a}{r_a} \, , \qquad
M=m_0+\sum_{a=1}^n \frac{m_a}{r_a} \, ,
\end{equation}

\noindent
(where the index $i$ takes three possible values $i=0,1,2$) and those in the non-Abelian sector

\begin{equation}
P=1+\sum_{a=1}^n \frac{\lambda_a}{r_a} \, , \qquad
Q=\sum_{a=1}^n \frac{\sigma_a \lambda_a}{r_a} \, , \qquad \text{with} \, \,  \lambda_a \geq 0 \, .
\end{equation}

\noindent
We refer to the $n$ poles of the harmonic functions as \emph{centers}, so $r_a\equiv \vert \vec{x} - \vec{x}_a \vert$ is the Euclidean distance from the $a$\textsuperscript{th} center. The first function, $H$, plays a special role. It determines the geometry of a four-dimensional Gibbons-Hawking ambipolar space \cite{Gibbons:2013tqa, Niehoff:2016gbi}, 

\begin{equation}
h_{mn} dx^m dx^n = H^{-1} (d\varphi +\chi)^{2}+H d\vec{x} \cdot d\vec{x} \, , \qquad \star_3 dH=d\chi \, .
\end{equation}

\noindent
To describe asymptotically flat solutions we need to recover four-dimensional Euclidean geometry in the base space in that limit. Hence, we shall impose $\sum_a q_a =1$. On the other hand, regularity at the centers demands that the Gibbons-Hawking charges $q_a$ are integer numbers, and therefore some of them must be negative.

The horizonless condition and the regularity of the metric at the centers require that the parameters $l^i_a$, $m_a$ and $\sigma_a$ are given by a certain combination of $q_a$, $k^i_a$ and other constants, which are specified in appendix~\ref{app:pelusos}. Moreover, asymptotic flatness also fixes the value of $m_0$ and imposes one constraint on the product of $l^0_0$, $l^1_0$ and $l^2_0$. The two remaining degrees of freedom in these constants are related to the moduli of the solution; that is, to the asymptotic value of the two Abelian scalars of the theory. Also, as discussed in the appendix, only $(n-1)$ of the $k^i_a$ parameters are \emph{physical}, as there is one redundant degree of freedom associated to gauge transformations of the vectors. 

Therefore, asymptotically flat horizonless configurations are specified by $4(n-1)$ \emph{charge} parameters ($k^i_a$ and $q_a$), the $2$ moduli parameters, the $n$ non-Abelian \emph{hair} parameters $\lambda_a$ and, of course, the coordinates of the centers, which add $3(n-2)$ degrees of freedom. In total, the parameter space of the solutions is $8(n-1)$-dimensional.

Not every point in the parameter space, however, yields a physically sensible solution. Actually, it is most frequent that a random election of such point gives a solution with closed timelike curves (CTCs) and Dirac-Misner string singularities connecting some of the centers. The absence of Dirac-Misner strings is achieved by imposing the so-called bubble equations,

\begin{equation}
\sum_{b\neq a} \frac{q_aq_b}{r_{ab}}\Pi^0_{ab}\left(\Pi^1_{ab} \Pi^{2}_{ab} - \frac{1}{2g^{2}}\mathbb T_{ab}\right)=\sum_{b,i} q_a q_b l_0^i  \Pi^i_{ab}.
\label{bbeq1}
\end{equation}

\noindent
where 

\begin{equation}
\Pi^i_{ab}=\frac{k_b^i}{q_b}-\frac{k_a^i}{q_a} \, , \qquad \mathbb T_{ab}=\frac{1}{q_a^2}+\frac{1}{q_b^2} \, ,
\end{equation}

\noindent
the non-Abelian gauge coupling constant is denoted by $g$ and $r_{ab}$ is the distance separating the centers $a$ and $b$. The combinations $\Pi^i_{ab}$ are the magnetic fluxes of the $i^{th}$ Abelian vector threading the non-contractible 2-cycle defined by the two centers $a$ and $b$. Notice that only $(n-1)$ equations are independent, as the sum in the index $a$ that labels the $n$ equations yields a trivial identity. Typically, solving the bubble equations constitutes a very hard step when constructing explicit microstate geometries. This is because, traditionally, those have been understood as equations for the variables $r_{ab}$, which have to be solved in terms of the independent charge parameters $k^i_a$ and $q_a$ and the moduli\footnote{The hair parameters $\lambda_a$ are absent in the bubble equations, although the non-Abelian fields are indirectly present through the term $\frac{1}{2g^{2}}\mathbb{T}_{ab}$.}. Then, after solving the system, usually by numerical methods, one finds that the obtained values of $r_{ab}$ rarely represent the distances between a collection of points, as they should all be real, positive numbers satisfying the triangle inequalities $r_{ac} \leq r_{ab} + r_{bc}$ for all $a$, $b$, $c$. To construct explicit microstate geometries, one usually relies on further restrictions that reduce the number of independent parameters but make it easier for the bubble equations to admit proper solutions.  

However, there seems to be no reason for considering the charge parameters more fundamental than the locations of the centers not only in the bubble equations, but also in the complete description of a particular microstate geometry. On one side, asymptotically the system \emph{looks like a black hole} and its main characteristics are determined by the charge parameters and the moduli. On the other side, the existence of well-separated centers is of the utmost importance for resolving the horizon, and their locations are responsible for the distinction between different microstates associated to the same black hole. In this paper we show that the bubble equations can be solved analytically in full generality, with complete access to the whole parameter space of regular solutions, by considering the location of the centers among the independent variables. 

As far as general CTCs are concerned, so far there has been no known analytically solvable restriction to guarantee their absence. Usually this has to be checked by evaluating numerically the positivity of the \emph{quartic invariant} of the solution, once the bubble equations have been solved and all parameters are already specified. In the next section we propose an algebraic condition on the space of parameters that allows us to distinguish whether or not the solution has CTCs, without making use of the numerical analysis of a function.

%However, not all the configurations of parameters that satisfy the bubbling equations are valid solutions. To avoid the global existence of closed timelike curves we need to satisfy the bound
%
%\begin{equation}
%\mathcal{I}_4 \equiv f^{-3} H - \omega_5^2 H^2 > 0 \, .
%\end{equation}
%We can choose the independent parameters such that this bound is satisfied in the centers and, afterwards, check numerically if the bound is satisfied in intermediate regions.

%%%%%%%%%%%%%%%%%%%%%%%%%%%%%%%%%%%%%%%%%%%%%%%%%%%%%%%%%%%%
%%%%%%%%%%%%%%%%%%%%%%%%%%%%%%%%%%%%%%%%%%%%%%%%%%%%%%%%%%%%
\section{The solution of the CTCs problem}
\label{sec:CTCproblem}

%%%%%%%%%%%%%%%%%%%%%%%%%%%%%%%%%%%%%%%%%%%%%%%%%%%%%%%%%%%%
\subsection{Solving the bubble equations analytically}
\label{solvbbeqs}

As we outlined in the previous section, the bubble equations have been traditionally solved for the distances between the centers using numerical methods. This, in turn, makes the task of constructing explicit microstate geometries complex and typically limits the regions of the parameter space that can be accessed. Here we use a different approach to address the problem that allows for the analytic resolution of the bubble equations in full generality. 

Since the number of independent magnetic fluxes associated to a given vector is the same as the number of independent equations, $(n-1)$, and those appear linearly in the system, it is reasonable to take them as the unknown variables for which the system is solved. As there are three different Abelian vectors, there are three possible ways in which we can write the system. In this section we write the explicit expressions when the $2$-fluxes are taken as the unknowns, although equivalent relations can be readily obtained for the $0$- and the $1$- fluxes. If we define

\begin{equation}
\alpha^{2}_{ab}=\frac{q_aq_b}{r_{ab}}\left(\Pi^0_{ab} \Pi^{1}_{ab}-l_0^2 r_{ab}\right) \, , \qquad \text{with} \, \, \alpha^2_{aa} = 0 \, ,
\end{equation}

\noindent
and

\begin{equation}
\beta^{2}_{a}=\sum_{\substack{b=1 \\ b \neq a}}^n \frac{q_aq_b}{r_{ab}}\left[ \frac{1}{2g^{2}}\mathbb T_{ab}\Pi^0_{ab}+\left(l_0^0 \Pi^0_{ab}+l_0^1 \Pi^1_{ab}\right)r_{ab}\right],
\end{equation}

\noindent
it is straightforward to see that (\ref{bbeq1}) can be rewritten as
\begin{equation}
\sum_{b=1}^n\alpha^{2}_{ab}\Pi^2_{ab}= \beta^{2}_a.
\label{bbeq2}
\end{equation}

This is a system of $n$ equations, but the sum of all of them is trivially satisfied\footnote{Notice that the fluxes $\Pi^i_{ab}$ are antisymmetric in their indices, while the matrix $\alpha^2_{ab}$ is defined to be symmetric. Also, we have that $\sum_a \beta^2_a=0$.}. Therefore, we can directly eliminate one of the equations, which is chosen to be the first one. We define the variables that will play the role of unknowns in the system of equations as follows

\begin{equation}
X^{2}_{\underline{a}}\equiv\Pi^{2}_{1( \underline{a}+1)}\, , \qquad  \underline{a}=1,\dots, n-1 \, .
\end{equation}

\noindent
Then, the rest of the 2-fluxes can be easily written in terms of these quantities as $\Pi^2_{(\underline{a}+1)(\underline{b}+1)}~=~X^{2}_{\underline{b}}-X^{2}_{\underline{a}}$. 

For this variables, we get a system of $(n-1)$ linear equations

\begin{equation}
\mathcal M^{2}_{\underline{a} \underline{b}}X^{2}_{\underline{b}}=B^{2}_{\underline{a}} \, , \qquad  %\underline{a} \, , \underline{b}=1,\dots, n-1 \, .
\label{eq:linearsystem}
\end{equation}

\noindent
where the components of the matrix $\mathcal M^{2}$ and the vector $B^{2}$ are given by 

\begin{equation}
\mathcal M^{2}_{\underline{a} \underline{b}}=\alpha^{2}_{(\underline{a}+1)(\underline{b}+1)}-\delta_{\underline{a}}^{\underline{b}}\sum_{c=1}^n \alpha^{2}_{(\underline{a}+1)c} \, , \qquad 
B^{2}_{\underline{a}}=\beta^{2}_{(\underline{a}+1)} \, .
\end{equation}

Thus, in our scheme the bubble equations can be solved by standard methods for an arbitrarily large number of centers\footnote{Provided computational resources are unlimited.}. This is, of course, if the solution exists. Let us go back some steps to understand this issue. We explained in the previous section that asymptotically flat horizonless configurations are determined by $8(n-1)$ parameters. To become regular microstate geometries, these configurations need to satisfy $(n-1)$ independent algebraic constraints known as bubble equations. This means that there are at most $7(n-1)$ independent parameters. In this section we have shown a way in which the independent parameters can be chosen in order to solve the bubble equations in full generality. But, still, there are special values of the independent parameters for which the bubble equations do not admit a solution: the values for which the determinant of the coefficient matrix $\mathcal M^2$ is zero.

To get some intuition about this, let us suppose for a moment that the $7(n-1)$ independent parameters are continuous variables. Then, the condition $\vert \mathcal M \vert = 0$ defines a codimension one hypersurface in the parameter space, to which we refer as a \emph{wall}. Although walls represent a very small region of the total space, we strongly believe that their presence is highly relevant. We plan to study this issue in detail in a forthcoming publication \cite{inprogress}.

%Let us put these ideas into practice. 

%%%%%%%%%%%%%%%%%%%%%%%%%%%%%%%%%%%%%%%%%%%%%%%%%%%%%%%%%%%%
\subsection{Absence of CTCs: an algebraic criterion}
\label{sec:CTC}

At this stage, there is one last restriction that physically sensible configurations must satisfy: the spacetime cannot contain closed timelike curves. As we already mentioned, this problem is translated to the positivity of a function, the quartic invariant

\begin{equation}
\label{eq:quartic}
\mathcal{I}_4 \equiv C^{IJK} Z_I Z_J Z_K H - \omega_5^2 H^2 \geq 0 \, ,
\end{equation}

\noindent
where we use the combinations (see the appendices for more information)

\begin{equation}
Z_I = L_I + 3 C_{IJK} \frac{\Phi^J \Phi^K}{H} \, , \qquad \omega_5 = M+\frac{1}{2} {L_I \Phi^{I}}{H^{-1}}+ C_{IJK} {\Phi^{I} \Phi^{J} \Phi^{K}}{H^{-2}} \, .
\end{equation}

The parameters are chosen such that $Z_I$ and $\omega_5$ do not diverge at the centers. Moreover, when the bubble equations are satisfied $\omega_5$ vanishes at the centers. Asymptotically, $Z_i$ (just the Abelian sector) go to the positive constant $l^i_0$ while $\omega_5$ goes to zero. In short, this means that $\mathcal{I}_4$ is dominated by the first factor both near the centers and far from them. Motivated by this observation, in this article we claim that the positivity of the quartic invariant is guaranteed if the first term is strictly positive,

\begin{equation}
\label{eq:CTCnew}
C^{IJK} Z_I Z_J Z_K H = Z_0 H (Z_1 Z_2 - \frac{1}{2} Z_\alpha Z_\alpha) > 0 \, ,
\end{equation}

\noindent
which implies that the term with $\omega_5$ is irrelevant for the study of CTCs, even at intermediate regions, unless the first term in $\mathcal{I}_4$ vanishes\footnote{While we have not been able to prove this claim in full generality, we have checked its validity in hundreds of thousands of pseudorandom configurations by computer analysis. In all the cases studied, the inequalities \eqref{eq:quartic} and \eqref{eq:CTCnew} are both true or both untrue.}. 

%Let us first consider the purely Abelian case ($Z_\alpha=0$). 

As $Z_i$ changes sign when $H$ does, the inequality \eqref{eq:CTCnew} can be converted into a collection of simpler inequations

\begin{equation}
\label{eq:CTCnew2}
Z_i H   > 0 \, , \qquad Z_1 Z_2 - \frac{1}{2} Z_\alpha Z_\alpha > 0 \, .
\end{equation}

\noindent
In terms of the parameters, these combinations of functions can be written as

\begin{eqnarray}
\nonumber
Z_i H &=& l^i_0 \sum_{a=1}^{n} \frac{q_a}{r_a} - 3 C_{ijk} \sum_{\substack{a,b=1 \\ a > b}}^{n} \frac{q_a q_b}{r_a r_b} \Pi^j_{ab} \Pi^k_{ab}
+\delta^0_i\frac{1}{2g^2} \sum_{a,b=1}^n \frac{1}{r_a r_b} \left( \frac{q_a}{q_b} - \frac{\lambda_a \lambda_b \vec{n}_a \cdot \vec{n}_b }{r_a r_b P^2} \right)
% =  \sum_{\substack{a,b=1 \\ a \geq b}}^{n} \frac{1}{r_a r_b} \left( l_0^i q_a r_b - 3 C_{ijk} q_a q_b \Pi^j_{ab} \Pi^k_{ab}  \right) 
\, , \\
Z_\alpha H &=& \sum_{\substack{a,b=1 \\ a \neq b}}^{n} \frac{q_a \lambda_b \Pi^0_{ab} } {g P r_a r_b^2} n^{(\alpha-2)}_b 
\, ,
\end{eqnarray}

\noindent
where $n^{(\alpha-2)}_a$ are the coordinates of the unit vector $\vec{n}_a \equiv \frac{\vec{x} - \vec{x}_a}{r_a} $ (recall that $\alpha=3,4,5$). Evaluating the Abelian functions $Z_i H$ at the centers we obtain

\begin{eqnarray}
\nonumber
\lim_{r_a \rightarrow 0} Z_0 H &=& \frac{1}{r_a} \left[ l^0_0 q_a -  \sum_{\substack{b=1 \\ b \neq a}}^{n} \frac{q_a q_b}{r_{ab}} \left( \Pi^1_{ab} \Pi^2_{ab} - \frac{1}{2g^2} \mathbb{T}_{ab} \right) +\frac{1}{g^2 \lambda_a} \left( \lambda_0+\sum_{b \neq a } \frac{\lambda_b}{r_{ab}} \right)  \right] + \mathcal{O}(r_a^0) \, , \\
\nonumber
\lim_{r_a \rightarrow 0} Z_1 H &=& \frac{1}{r_a} \left[ l^1_0 q_a -  \sum_{\substack{b=1 \\ b \neq a}}^{n} \frac{q_a q_b}{r_{ab}}  \Pi^0_{ab} \Pi^2_{ab}  \right] + \mathcal{O}(r_a^0) \, , \\
\lim_{r_a \rightarrow 0} Z_2 H &=& \frac{1}{r_a} \left[ l^2_0 q_a -\sum_{\substack{b=1 \\ b \neq a}}^{n} \frac{q_a q_b}{r_{ab}}  \Pi^0_{ab} \Pi^1_{ab}  \right] + \mathcal{O}(r_a^0) \, ,
\end{eqnarray}

\noindent
and, from the first set of inequalities in \eqref{eq:CTCnew2}, we find that the combination of parameters inside the brackets must be positive for all centers. Notice that in these expressions the purely Abelian limit is effectively recovered by taking $g \rightarrow \infty$, and that in this limit the last inequality in \eqref{eq:CTCnew2} is trivial. At first sight, it is noteworthy that these combinations of parameters look very similar to the elements in the diagonal of the coefficient matrices $\mathcal M^0$, $\mathcal M^1$ and $\mathcal M^2$, which are

\begin{equation}
\mathcal M^i_{(a-1)(a-1) }= -\sum_{\substack{b=1 \\ b \neq a}}^{n} \frac{q_a q_b}{r_{ab}} \left(3 C_{ijk} \Pi^j_{ab} \Pi^k_{ab} - \delta^i_0 \frac{1}{2g^2} \mathbb{T}_{ab} \right) +   l^i_0 q_a (1-q_a) \, .
\end{equation}

\noindent
But, however, what it is truly remarkable is that the positivity of the elements in the diagonal of those matrices is sufficient to ensure the positivity of the divergences of the functions $Z_i H$ at the centers\footnote{Except, perhaps, at the first center, for which the coefficient of the divergence is not directly related to any diagonal element of the coefficient matrices $\mathcal M^i$. This is because in the previous section we decided to eliminate the first of the bubble equations and take $\Pi^i_{1b}$ as the unknowns. Of course, it is possible to take any other center as reference, obtaining additional conditions to guarantee the positivity of the divergence at the first center.}. This suggests that there might be a relation between the properties of the linear system of bubble equations and the absence of CTCs. To understand this relation better, it is convenient to consider simple configurations.

% Moreover, studying the divergence at the centers of the second inequality in \eqref{eq:CTCnew2} we get
%
%\begin{equation}
%\left( l^1_0 q_a -  \sum_{\substack{b=1 \\ b \neq a}}^{n} \frac{q_a q_b}{r_{ab}}  \Pi^0_{ab} \Pi^2_{ab}  \right)
%\left( l^2_0 q_a -\sum_{\substack{b=1 \\ b \neq a}}^{n} \frac{q_a q_b}{r_{ab}}  \Pi^0_{ab} \Pi^1_{ab}  \right)
%-\left( \sum_{\substack{b=1 \\ b \neq a}}^{n} \frac{q_b \Pi^0_{ab}}{g r_{ab} } \right)^2 > 0 \, .
%\end{equation}
%
%\noindent
%Therefore, we have obtained a large set of necessary but, in principle, not sufficient conditions to avoid the presence of CTCs. 

The study of two-center, purely Abelian microstate geometries provides a great deal of insight in this problem. In this case the bubble equation is

\begin{equation}
\label{eq:bubeq2cen}
-\frac{q_1 q_2}{r_{12}} \left( \Pi^0_{12} \Pi^1_{12} - l^2_0 r_{12} \right) X^{2}_1 = - q_1 q_2 \left( l^0_0 \Pi^0_{12} + l^1_0 \Pi^1_{12} \right) \, ,
\end{equation}

\noindent
where $X^{2}_1 = \Pi^2_{12}$. This configuration will not contain CTCs if

\begin{equation}
Z_i H = \frac{1}{r_1 r_2} \left[ l_0^i \left( q_1 r_2 + q_2 r_1 \right) - 3 C_{ijk} q_1 q_2 \Pi^j_{12} \Pi^k_{12} \   \right] >0  \, ,
\end{equation}

\noindent
for $i=0,1,2$. Since $r_1$ and $r_2$ are positive numbers, we just need to impose the positivity of the function inside the bracket. Without loss of generality, we can take $q_1$ to be positive and $q_2$ to be negative, with $q_1+q_2=1$. Then, the function $(q_1 r_2 + q_2 r_1)$ is bounded from below by the number $q_2 r_{12}$, so we can write

\begin{equation}
Z_i H \geq \frac{1}{r_1 r_2} \left( l_0^i  q_2 r_{12} - 3 C_{ijk} q_1 q_2 \Pi^j_{12} \Pi^k_{12} \   \right)  \, .
\end{equation}

\noindent
In particular, for $Z_2 H$ we need the combination $\left( l_0^2  q_2 r_{12} - q_1 q_2 \Pi^0_{12} \Pi^1_{12} \   \right) $ to be positive. This combination looks very similar to the eigenvalue of the matrix $\mathcal M^2$ (in this case a $1 \times 1$ matrix) that defines the bubble equation \eqref{eq:bubeq2cen}. Actually, as $q_1 q_2 < q_2$, if the eigenvalue is positive then the above combination is positive and therefore $Z_2 H>0$. 

In purely Abelian configurations, the three different type of fluxes appear in the bubble equations exactly in the same manner. Then, the bubble equations can be solved for the $0$-fluxes or the $1$-fluxes, defining two more matrices $\mathcal M^{0}$ and $\mathcal M^{1}$ respectively. Following the same reasoning as in the previous paragraph, we conclude that if the eigenvalues of the three matrices $\mathcal M^{i}$ are positive then $Z_i H>0$ and therefore the configuration is free of CTCs. Of course, $\mathcal M^{0}$ and $\mathcal M^{1}$ depend on the $2$-fluxes, which play the role of unknowns in the bubble equations. So, naively, it might seem that this criterion for identifying CTCs is of little help in practice, as we would like to dispose of a set of conditions on the parameter space only. However, using the bubble equations, it is immediate to prove that the eigenvalues of the three matrices are of the same sign. For example, for $\mathcal M^{0}$ we have that its eigenvalue is positive if $ \Pi^1_{12} \Pi^2_{12} - l^0_0 r_{12}>0$. This combination can be rewritten using the bubble equations (\ref{eq:bubeq2cen}) as

\begin{equation}
 \Pi^1_{12} \Pi^2_{12} - l^0_0 r_{12}  = 
 %\Pi^1_{12} \frac{ \left( l^0_0 \Pi^0_{12} + l^1_0 \Pi^1_{12} \right) r_{12} }{\Pi^0_{12} \Pi^1_{12} - l^2_0 r_{12}} - l^0_0 r_{12} = 
 \frac{ l^1_0 (\Pi^1_{12} )^2 r_{12} + l^0_0 l_0^2 (r_{12})^2  }  {\Pi^0_{12} \Pi^1_{12} - l^2_0 r_{12}} \, .
\end{equation}

\noindent
Since the numerator on the right hand side is positive, the left hand side has the same sign as the denominator, which proves our point.

At this stage we have shown that if the eigenvalue of $\mathcal M^{2}$ is positive, then we have $Z_i H>0$ and, according to our claim at the beginning of this section, the quartic invariant is positive and the solution does not contain CTCs. But, what would happen if the eigenvalue were negative? According to the preceding discussion it might be possible to have $Z_2 H > 0 $ even when the eigenvalue is not positive. That is, it is possible to choose the parameters such that

\begin{equation}
\left( l_0^2  q_2 r_{12} - q_1 q_2 \Pi^0_{12} \Pi^1_{12} \   \right)  > 0 > \left( l_0^2 q_1  q_2 r_{12} - q_1 q_2 \Pi^0_{12} \Pi^1_{12} \   \right) \, .
\end{equation}

\noindent
Remarkably, it turns out that these inequalities imply that $Z_0 H$ and $Z_1 H$ eventually become negative! For instance, we can check it explicitly for the latter (both proofs are identical). In first place, notice that the first inequality requires $\Pi^0_{12} \Pi^1_{12} > 0 $. Using the bubble equations we can write

\begin{equation}
Z_1 H = \frac{l^0_0\left[ -q_1 q_2 r_{12} \left( \Pi^0_{12} \right)^2 +  \left(\Pi^0_{12} \Pi^1_{12} - l^2_0 r_{12} \right) \left( q_1 r_2 + q_2 r_1 \right) - \Pi^0_{12} \Pi^1_{12} q_1 q_2 r_{12} \right] }{r_1 r_2 \left(\Pi^0_{12} \Pi^1_{12} - l^2_0 r_{12} \right) } \, .
\end{equation}

\noindent
This function is positive asymptotically, but negative at $r_2=0$, $r_1=r_{12}$ whenever $\left(\Pi^0_{12} \Pi^1_{12} - l^2_0 r_{12} \right) < 0$. 

In summary, we have proved that Abelian two-center microstate geometries do not contain CTCs if and only if the eigenvalue of $\mathcal M^{2}$ is positive. The same conclusion can be obtained for non-Abelian two-center microstate geometries. In this case the proof is similar, although it is more technical and not particularly illuminating. In view of this result and based on the observations exposed at the beginning of this section for multicenter configurations, we make the following proposal.

\textbf{Conjecture:} Five-dimensional microstate geometries on a Gibbons-Hawking base, with or without non-Abelian fields, do not contain CTCs if and only if the coefficient matrix of the bubble equations is positive-definite. 

If our conjecture is true, the construction of five-dimensional microstate geometries without CTCs will no longer require the numerical evaluation of any function on $\mathbb{R}^3$, but it will be sufficient to check an algebraic property of a matrix. This would extraordinarily simplify the problem of describing and studying this type of supergravity solutions, giving rise to a new plethora of smooth geometries.

To close this section let us mention that, although we have not been able to prove our conjecture in full generality, we have tested its validity with a large number of multicenter configurations. We have analyzed more than 100,000 solutions with pseudo-random parameters, finding a perfect agreement with our proposal.

%%%%%%%%%%%%%%%%%%%%%%%%%%%%%%%%%%%%%%%%%%%%%%%%%%%%%%%%%%%%
%%%%%%%%%%%%%%%%%%%%%%%%%%%%%%%%%%%%%%%%%%%%%%%%%%%%%%%%%%%%

\subsection{Contractible clusters and scaling solutions}
\label{sec:scaling}

Scaling microstate geometries can be defined as solutions for which the centers can be brought arbitrarily close without significantly modifying the asymptotic charges, \cite{Bena:2006kb}. As the centers approach each other, the geometry of the system does not only reproduce the asymptotic charges of an extremal black hole, but also starts to \emph{look like one} at intermediate regions. In the zero-size limit all centers merge, a horizon is developed and the configuration becomes a black hole. But right before reaching the black hole limit, the solution is still horizonless and partially reproduces the \emph{throat} that characterizes the near-horizon geometry of a black hole, capping off smoothly at some finite depth, although arbitrarily large. It is for this reason that scaling microstate geometries are expected to correspond to the classical description of individual microstates of a black hole, \cite{Bena:2007kg}. We now show that the formalism we have presented is extraordinarily well-suited to describe and study scaling solutions. 

We consider scaling solutions that preserve the shape of the distribution\footnote{Ideal scaling solutions would preserve the asymptotic charges while slightly modifying the relative distances of the cluster. However these type of scalings are extremely hard to describe and, as we are more interested in the \emph{scaled} configurations than in the \emph{scaling} process itself, we ignore this issue.}. This means that we can write the distances between centers as

\begin{equation}
r_{ab} = \mu d_{ab} \, ,
\label{eq:scalingtrans}
\end{equation}

\noindent
where $d_{ab}$ remain constant in the scaling process, which is controlled by varying $\mu$ to arbitrarily small positive numbers. We can define the following quantities,

\begin{equation}
\bar{\alpha}^2_{ab}=\frac{q_aq_b}{d_{ab}}\Pi^0_{ab} \Pi^{1}_{ab} \, , \qquad
\mathring{\alpha}^2_{ab}=-q_a q_b l_0^2 \, ,
\end{equation}

\noindent
and

\begin{equation}
\bar\beta^2_{a}=\sum_{b=1}^n \frac{q_a q_b}{d_{ab}} \frac{4}{g^{2}}\mathbb T_{ab}\Pi^0_{ab} \, , \qquad
\mathring\beta^2_{a}=\sum_{b=1}^n q_a q_b \left(l_0^0 \Pi^0_{ab}+l_0^1 \Pi^1_{ab}\right) \, ,
\end{equation}

\noindent
which are manifestly invariant during the scaling process. Then, upon substitution of \eqref{eq:scalingtrans} in \eqref{eq:linearsystem}, the bubble equations can be written as

\begin{equation}
\left(\bar{\mathcal M}^2_{\underline{a}\underline{b}}+ \mu \mathring{\mathcal M}^2_{\underline{ab}} \right) X^2_{\underline{b}}=\bar B^2_{\underline{a}} + \mu \mathring B^2_{\underline{a}} \, .
\label{eq:linearsystemmu}
\end{equation}

\noindent
If compared with the original equation, we have $\mu \mathcal M^2 =\bar{\mathcal M}^2 + \mu \mathring{\mathcal M}^2$ for the coefficient matrix and $\mu B^2 = \bar B^2 + \mu \mathring B^2$ for the column vector. In terms of the parameters, we have

\begin{eqnarray}
\bar{\mathcal M}^{2}_{\underline{a} \underline{b}}=\bar \alpha^{2}_{(\underline{a}+1)(\underline{b}+1)}-\delta_{\underline{a}}^{\underline{b}}\sum_{c=1}^n \bar \alpha^{2}_{(\underline{a}+1)c} \, , \qquad 
\bar B^{2}_{\underline{a}}= \bar \beta^{2}_{(\underline{a}+1)} \, , \\
\mathring{\mathcal M}^{2}_{\underline{a} \underline{b}}=\mathring \alpha^{2}_{(\underline{a}+1)(\underline{b}+1)}-\delta_{\underline{a}}^{\underline{b}}\sum_{c=1}^n \mathring \alpha^{2}_{(\underline{a}+1)c} \, , \qquad 
\mathring B^{2}_{\underline{a}}= \mathring \beta^{2}_{(\underline{a}+1)} \, .
\end{eqnarray}

The bubble equations as written in \eqref{eq:linearsystemmu} are well defined even in the zero-size limit $\mu = 0$, where they cease to have a physical meaning. Scaling solutions can be identified as those for which one can take the zero-size limit through a continuous transformation and still obtain a valid solution of the bubble equations, without any of the asymptotic charges becoming zero. The existence of this limit cannot be taken for granted. Actually, for purely Abelian solutions one always has $\bar B^2=0$ and the bubble equations become a homogeneous system in the zero-size limit. Then, there are non-trivial solutions (that is, solutions with $X^2_{\underline{a}} \neq 0$ for some $\underline{a}$) only if the determinant of the corresponding coefficient matrix, $\bar{\mathcal M}^2$, vanishes. In other words, purely Abelian scaling solutions necessarily flow to special points of the parameter space when taking the zero-size limit. It is for this reason that many Abelian solutions cannot be scaled without some of the asymptotic charges becoming zero. 

The situation is completely different when non-Abelian fields are also considered. In this situation we have $\bar B^2 \neq 0$ and the system is still inhomogeneous in the zero-size limit. This means that non-Abelian microstate geometries can \emph{typically} be scaled\footnote{We will study this issue in a forthcoming publication. The reason why some solutions might not be scaling is because they need to cross a \emph{wall} in the parameter space while being subjected to the scaling process. On the other hand, since non-Abelian configurations can always be truncated to Abelian solutions, we can always recover Abelian scaling configurations if the truncation is implemented at some stage of the scaling process.}. Actually, from our point of view this is the most important contribution that non-Abelian fields bring to the ``microstate geometries program''. Typically these fields enter the solutions modifying the spacetime metric, the asymptotic charges and the size of the bubbles very softly; in most cases these physical properties are practically preserved after introducing the non-Abelian distortion. A reason for it is that, from the $10$-dimensional Heterotic supergravity perspective, these fields appear at first order in $\alpha ' $. However, this distortion becomes critical when we take the zero-size limit, enlarging the spectrum of scaling solutions.

\section{One Thousand and One Bubbles}
\label{sec:thousand}

%%%%%%%%%%%%%%%%%%%%%%%%%%%%%%%%%%%%%%%%%%%%%%%%%%%%%%%%%%%%
\subsection{Exact solutions on lines and circles}
\label{sec:normal_axi}

There is one issue that might worry some of the readers: the fluxes that solve the equations are in general irrational numbers. This is because the distances between a collection of points in three dimensions are usually irrational. The asymptotic charges, which are expected to be quantized when these solutions are properly interpreted within the context of string theory, are directly related to the fluxes, see equations \eqref{eq:charges}. Therefore, we would prefer the fluxes to be given by, at least, rational quantities. This fact can be seen as a motivation for the traditional approach to solve the bubble equations, in which the fluxes are guaranteed to be rational. 

However, as the method we present here is completely general, it must also describe all microstate geometries with rational fluxes. These classes of solutions are obtained if the centers are chosen such that all the relative distances between them are rational numbers. An obvious possibility is to take all centers laying on a line, so the solution is axisymmetric. This kind of microstate geometries are easier to build and study, and most of the explicit constructions known are axisymmetric. As a first application of the aforementioned procedure we show in Table \ref{tab:exact} a first example of a $5$-center solution\footnote{Our criterion to find a solution free of CTCs is to systematically look for parameters for which the coefficient matrix $M^2$ is positive definite. In any case, we have also checked numerically the absence of CTCs for all the examples displayed in this article.}. Motivated by the recent results of \cite{Heidmann:2017cxt, Bena:2017fvm}, the locations of the centers present a \emph{hierarchical structure}; i.e. the values of the relative distances vary between different orders of magnitude. As argued in those articles, such structure potentially favors finding solutions whose angular momentum is far from maximal. The $5$ center solution described here has negligible angular momentum, see Table~\ref{tab:normal_momenta}, and constitutes the first five-dimensional microstate geometry that exhibits this property. As shown in Table \ref{tab:exact}, the solution can be scaled without any problem introducing a conformal factor $\mu$ for the coordinates of the centers.

\begin{table}[phtb]
  \centering
  %\sffamily
  \scriptsize
  \onehalfspacing
  \captionof{table}{\small Input and output parameters of solutions. $l^1_0=\sqrt{2}$, $l^2_0=1/\sqrt{2}$ and $g=1$ for all the cases. Output parameter values shown are only approximate.}
  \label{tab:exact}
  
 % \vspace{-0.2em}
 
     \begin{tabular}{*{11}{c}}
    \toprule
    \multicolumn{6}{c}{\textbf{5 centers on a line}}\\
    \midrule
       $x$&-1& -0.999& 0& 0.999& 1\\\midrule
       $q$&1& -1& 1& -1& 1\\
       $k^0$&10& -27& -37& 17& 1\\
       $k^1$&38& -68& 46& 14& -11\\\midrule
       $k^2(\mu=1)$&1& -1.03739& 0.223899& 2.25387& -1.79766\\
       $k^2(\mu=0.0005)$&1& -1.03707& 0.2834& 2.10992& -1.65908\\
    \bottomrule\\
    \end{tabular}%
   %   \vspace{-0.2em}

    \begin{tabular}{*{11}{@{ }c@{ }}}
    \toprule
    \multicolumn{11}{c}{\textbf{10 centers on a circle}}\\\midrule
    \multicolumn{11}{c}{$l_1=\left(\frac{2001}{2002001}\right)^2$, $h_1=2001\cdot\frac{2002000}{2002001^2}$, $l_2=\left(\frac{6001}{18006001}\right)^2$, $h_2=6001\cdot\frac{18006000}{18006001^2}$}\\
    \midrule
       $x$&0.5& $0.5 -l_2$& $0.5 -l_1$&$-0.5 + l_1$& $-0.5 +l_2$& -0.5& -0.5 + $l_2$& $-0.5 +l_1$&$0.5 - l_1$& $0.5 -l_2$\\
       $y$&0& $h_2$ & $h_1$ & $h_1$ &$h_2$ & 0& $-h_2$&$-h_1$&$-h_1$ & $-h_2$\\\midrule
       $q$&2& -1& 1& -1& 1& -1& 1& -1& 1& -1\\
       $k^0$&32& 72& 12& 60& 39& 30& 38& 11& 9& 51\\
       $k^1$&51& 99& 32& 24& 90& 11& 57& 26& 9& 78\\\midrule
       $k^2(\mu=1)$&1& -0.495548& 0.503203& -0.461338& 0.467769& -0.456857& 0.471085&-0.454882& 0.505734& -0.493379\\
       $k^2(\mu=0.0005)$&1& -0.495561& 0.503179& -0.483632& 0.490029& -0.479143& 0.493333&-0.477239& 0.505681& -0.493401\\
    \bottomrule\\
    \end{tabular}
    
   % \vspace{-0.2em}
    
    \begin{tabular}{*{7}{@{ }c@{ }}}
    \toprule
    \multicolumn{7}{c}{\textbf{6 centers on a circle}}\\\midrule
    \multicolumn{7}{c}{$l=\left(\frac{2001}{2002001}\right)^2$, $h=2001\cdot\frac{2002000}{2002001^2}$}\\
    \midrule
       $x$&0.5& $0.5 - l$ & $l - 0.5$ & -0.5 & $l - 0.5$ & $0.5-l$\\
       $y$&0&\it h&\it h& 0& \it -h&\it -h\\\midrule
       $q$&2& -1& 1& -1& 1& -1\\
       $k^0$&-100& 69& 46& -95& -7& 73\\
       $k^1$&-98& 56& -15& -68& 36& 79\\\midrule
       $k^2(\mu=1)$&1& 0.133637& 11.6034& -11.6405& 11.598& -0.421436\\
       $k^2(\mu=0.0005)$&1& 0.102875& 10.9491& -10.9852& 10.9439& -0.425198\\
    \bottomrule\\
    \end{tabular}%
  
\end{table}%

In order to go beyond axisymmetry, we now define a very interesting, arbitrarily large set of points with rational relative distances lying on a circle. The result is based on the original proof of the Erd\H os-Anning theorem, \cite{anning1945}, that states that any infinite collection of points can have integral distances only if these are aligned. However, as we are about to see, it is possible to have an infinite set of points with mutual rational distances. First, pick a circle with unit diameter centered at the origin of coordinates. A primitive Pythagorean triple\footnote{Primitive Pythagorean triples are generated through Euclid's formula,
\begin{equation}
a_i=m_i^2-n_i^2 \, , \quad b_i=2 m_i n_i \, , \quad c_i= m_i^2+n_i^2\, ,
\end{equation}
for any pair of coprime integers $m_i > n_i >0 $.} $P_i$ is composed of three coprime natural numbers $a_i$, $b_i$ and $c_i$ such that $a_i^2+b_i^2=c_i^2$. The triple $P_i$ defines a right triangle whose hypotenuse and catheti lengths are $1$, $a_i/c_i$ and $b_i/c_i$ respectively. This triangle can be place such that the hypotenuse lies on the x-axis and the coordinates of the vertices are $(-\frac{1}{2},0)$, $(-\frac{1}{2}+l_i, h_i)$ and $(\frac{1}{2},0)$, where $l_i=\left(\frac{a_i}{c_i}\right)^2$ and $h_i=\frac{a_i b_i}{c_i^2}$. Then, the triangle defines three points at rational distances on the unit diameter circle. 

In virtue of Ptolemy's theorem, any other point with rational distances to the pair of points $(-\frac{1}{2},0)$ and $(\frac{1}{2},0)$ is necessarily separated by a rational distance from $(-\frac{1}{2}+l_i, h_i)$ as well. In particular, this means that we can use the same Pythagorean triple $P_i$ to find three more valid points: $(-\frac{1}{2}+l_i, -h_i)$, $(\frac{1}{2}-l_i, h_i)$ and $(\frac{1}{2}-l_i, -h_i)$. Any additional primitive triple can add up to four points more to the set in the obvious manner. Moreover, since the four points associated to a triple define two new diameters of the circle those can also be used as hypotenuses, providing new possibilities to enlarge the collection. The procedure can be prolonged without end, defining a dense set of points on the circle. Finally, the value of the radius can be set to any rational number $\mu$. Therefore, these configurations are very well-suited to build scaling solutions. 

Table \ref{tab:exact} contains a couple of examples of microstate geometries with $6$ and $10$ centers lying on a circle. Once again, a hierarchic structure has been imposed by making use of Pythagorean triples for which $a_i \ll b_i$. 
% We have checked that the solutions are free of CTCs by numerical inspection\footnote{As these configurations are not axisymmetric, it is necessary to check the positivity of the quartic invariant $\mathcal{I}$ not only at the plane that contains the centers, but also we need to study its evolution as we move far away from that plane.}

\begin{table}[phtb]
  \centering
  \scriptsize
  \singlespacing
  \captionof{table}{\small Input and output parameters of a 50 centre example. $l^1_0=\sqrt{2}$, $l^2_0=1/\sqrt{2}$ and $g=1$. Output parameter values shown are only approximate.}
 % \label{tab:exact}
    \begin{tabular}{*{5}{c}}
     \toprule
    \multicolumn{5}{c}{\textbf{50 centers on a line}}\\
    \midrule
    $x$&$q$&$k^0$&$k^1$&$k^2$\\\midrule
    0.0330053&2&-20&-55&1\\
    0.0984265&-1&-32&-14&-0.540293\\
    -0.0179676&1&-70&-52&0.510139\\
    0.092019&-1&-33&-2&-0.544646\\
    0.011303&1&-33&-56&0.507942\\
    0.0159932&-1&-42&-97&-0.513755\\
    -0.0419008&1&-59&-83&0.506483\\
    0.00449896&-1&-30&-35&-0.523255\\
    -0.0371543&1&-83&-82&0.520376\\
    0.0249915&-1&-13&-61&-0.523077\\
    0.904343&1&-66&-90&0.51483\\
    0.966033&-1&-100&-27&-0.521793\\
    1.06745&1&-83&-11&0.500632\\
    0.991016&-1&-40&-89&-0.518794\\
    0.918601&1&-79&-38&0.507973\\
    1.09964&-1&-27&-65&-0.515174\\
    0.998465&1&-17&-28&0.502238\\
    0.913144&-1&-41&-12&-0.529991\\
    1.09778&1&-12&-31&0.50078\\
    0.959097&-1&-99&-71&-0.515806\\
    2.04383&1&-74&-77&0.531604\\
    2.03968&-1&-6&-7&-0.561911\\
    1.92718&1&-95&-77&0.531914\\
    1.97688&-1&-23&-78&-0.52514\\
    1.90891&1&-95&-33&0.509497\\
    1.98718&-1&-74&-13&-0.525948\\
    1.95846&1&0&-37&0.446919\\
    2.03144&-1&-46&-7&-0.541284\\
    1.99207&1&-53&-25&0.50149\\
    2.04206&-1&-57&-3&-0.540974\\
    2.96983&1&-9&-84&0.500777\\
    2.92655&-1&-54&-27&-0.515791\\
    2.94343&1&-1&0&0.430675\\
    2.96789&-1&-59&-35&-0.514717\\
    2.96737&1&-55&-7&0.49317\\
    2.97213&-1&-83&-25&-0.511543\\
    2.99724&1&-17&-42&0.49296\\
    2.93693&-1&-1&-61&-0.536688\\
    2.99367&1&-19&-15&0.482956\\
    3.09503&-1&-70&-48&-0.515729\\
    3.97904&1&-35&-38&0.526046\\
    3.98217&-1&-85&-11&-0.540595\\
    4.09649&1&-46&-39&0.491727\\
    3.99963&-1&-17&-2&-1.11207\\
    4.0424&1&-78&-63&0.506147\\
    4.03426&-1&-67&-54&-0.497783\\
    3.99249&1&-51&-20&0.556606\\
    4.01874&-1&-72&-93&-0.452374\\
    4.02437&1&-30&0&0.443444\\
    4.0978&-1&-61&-80&-0.499799\\\bottomrule
    \end{tabular}
  \label{tab:50_centre_solutions}
  
\end{table}%

Some of these solutions have more bubbles than any previously known example and, furthermore, can be specified with exact accuracy. Increasing the number of centers is feasible, although computationally demanding. On one side, solving a linear system of equations is a problem of complexity P of order $\mathcal{O}(n^3)$. That is, the time required to solve the bubble equations approximately scales with the cube of the number of centers. On the other side, increasing the number of centers seems to favor the appearance of CTCs, so it is more likely that random elections of the parameters yield to unphysical solutions. According to our conjecture, this is just a natural consequence; as the coefficient matrix becomes bigger it is harder and harder to find the parameters such that all its eigenvalues are positive. Nevertheless, we have been able to describe solutions with a very large number of centers by focusing on regions of the parameter space that seem to favor the coefficient matrix is positive definite\footnote{Our main guides are to impose the presence of hierarchical structures and to take all $k^{0,1}_a$ coefficients of the same sign.}. A particular example of an axisymmetric $50$-center solution is given in Table \ref{tab:50_centre_solutions}.

%The asymptotic charges, angular momentum and entropy parameter of these configurations are given in Table~\ref{tab:normal_momenta}.

Another interesting issue is that the parameters $\lambda_a$ that determine the non-Abelian seed functions seems to play a subleading role in the CTCs problem. In fact, once a solution without CTCs has been found, we can generate as many as we want by modifying the non-Abelian parameters\footnote{We have checked this by taking arbitrary values of the non-Abelian parameters in a finite range. However, based on how these parameters appear in the non-Abelian seed functions, we think that one can take any positive value for them and CTCs will not appear.}, as long as all of them remain positive. This is the reason why we have not specified any particular values in the tables. Therefore, the inclusion of non-Abelian fields not only makes it easier to find scaling solutions, but also enlarges the number of solutions with a given set of asymptotic charges, as expected \cite{Ramirez:2016tqc}.

%
%
%
%Asymptotic charges and angular momentum of these solutions have also been calculated. Their values are gathered up in Table~\ref{tab:normal_momenta}. Most of the solutions reported so far in the literature are maximally spinning, id est, the quotient $\mathcal{I}$ takes values about $\sim 99\%$. In spite of hierarchy, this is the case of three of the examples shown here, although their quotients are not so close to 1 as usual. However, the 6-center circle solution provides an example of non-maximal spinning, quite a remarkable result since its quotient value is even lower than other examples of the same kind.
%

\begin{table}
\centering
\onehalfspacing
\scriptsize
\begin{tabular}{*{5}{c}}
\toprule
&\bf 50-center line&\bf 5-center line&\bf 10-center circle&\bf 6-center circle\\\midrule
$\mathcal{Q}_0$&2378.38&46.7351&48.5981&448.285\\
$\mathcal{Q}_1$&2749.4&43.2695&37.403&509.907\\
$\mathcal{Q}_2$&5058525&2723&175063&12273\\
$J_R$&-$5.7506\cdot10^6$&-1.0708&17689.5&25956.7\\\midrule
$\mathcal{H}$&$2.7\cdot10^{-4}$&$0.9999998$&0.017&0.76\\\bottomrule
\end{tabular}

\captionof{table}{Asymptotic charges and angular momenta of the solutions for $\mu=1$.}
\label{tab:normal_momenta}
\end{table}

%%%%%%%%%%%%%%%%%%%%%%%%%%%%%%%%%%%%%%%%%%%%%%%%%%%%%%%%%%%%

\subsection{General locations}

It is comforting that we can use our method to describe, for the first time, many-center five-dimensional microstate geometries with an exact accuracy. Nevertheless, we are also interested in the possibility of describing more general solutions with centers at general locations, which includes the possibility of having irrational distances. In practice, this implies that the bubble equations must be solved approximately. We distinguish two possibilities:

\begin{itemize}
\item \textbf{Approximate fluxes}. The first possibility is to solve the bubble equations for the fluxes. In this case these will be given by irrational numbers and, as discussed at the beginning of the preceding subsection, this can be considered inconvenient because they are related to the asymptotic charges. Then, one valid option is to round the fluxes such that the charges take valid values, and admit that the solution is only specified approximately. This can be a useful possibility when one is interested in studying generic properties of the solutions, rather than in performing a very precise analysis.

\item \textbf{Approximate locations}. The procedure that we follow to avoid having approximate fluxes can be summarized as follows. In a first step, we choose our favorite distribution of centers and solve the bubble equations for the fluxes. Then, we round the values and solve again the equations for the distances between the centers, using the fluxes as input data now. We expect the distances not to change too much for small enough changes of the fluxes. 
Once we know the distances, we have to place the centers in the tridimensional space $\mathbb R^3$. Unfortunately, this can only be done in full generality for four centers at most, so in configurations with more centers one has to impose restrictions in the locations when solving the bubble equations the second time (for example, one can consider axisymmetric configurations only).

\end{itemize}

%Restricting ourselves to four-center distributions, the positions of the centers can be found as follows. If we placed the first two points along the $x$-axis, points 3 and 4 must be in circles of radius $R_3$ and $R_4$ that are contained in the planes $x=\underline x_3$ and $x=\underline x_4$, respectively. If the first point is at the origin $\vec x_1=\left(0,0,0\right)$, the second must be therefore at $\vec x_2=\left(r_{12},0,0\right)$ and one gets that
%
%\begin{eqnarray}
%\underline x_3=\frac{r_{13}^2+r_{12}^2-r_{23}^2}{2r_{12}},\\
%\underline x_4=\frac{r_{14}^2+r_{12}^2-r_{24}^2}{2r_{12}},
%\end{eqnarray}
%and 
%
%\begin{eqnarray}
%R_3^2=r_{13}^2-\underline x_3^2,\\
%R_4^2=r_{14}^2-\underline x_4^2.
%\end{eqnarray}
%Then, we have that the coordinates of points 3 and 4 can be specified as 
%
%\begin{eqnarray}
%\vec x_3=\left(\underline x_3, R_3\cos \varphi_3, R_3\sin\varphi_3\right),\\
%\vec x_4=\left(\underline x_4, R_4\cos \varphi_4, R_4\sin\varphi_4\right),\\
%\end{eqnarray}
%for some angles $\varphi_3$ and $\varphi_4$. Using the fact that the distance between these points is $r_{34}$, one finds that 
%both are related through
%\begin{equation}
%\varphi_3-\varphi_4=\pm\arccos \left[\frac{R_3^2+R_4^2+\left(\underline x_3-\underline x_4\right)^2-r_{34}^2}{2R_3 R_4}\right].
%\end{equation}

%%%%%%%%%%%%%%%%%%%%%%%%%%%%%%%%%%%%%%%%%%%%%%%%%%%%%%%%%%%%%%%%
%%%%%%%%%%%%%%%%%%%%%%%%%%%%%%%%%%%%%%%%%%%%%%%%%%%%%%%%%%%%%%%%
\section{Final comments and further directions}
\label{sec:conclusions}

In this article we have presented an efficient method to construct general five-dimensional supersymmetric microstate geometries on a Gibbons-Hawking base. We have conjectured that the CTCs problem can be solved through the evaluation of a simple algebraic relation without the need to numerically evaluate the quartic invariant function. We have accompanied the exposition with a few explicit solutions, which were found making use of our method. These solutions exhibit novel properties in their class, such as arbitrarily small angular momentum or large number of centers, being some of them not axisymmetric distributions. This not only reveals that the spectrum of smooth microstate geometries on a Gibbons-Hawking base is actually very rich, but also that it is possible to find and study this type of solutions.

In particular, this method can be used to describe simple five-dimensional smooth, horizonless scaling solutions with the asymptotic charges of a D1-D5-P black hole without angular momentum. It would be interesting to study general properties of these geometries and compare them with those of a black hole; their geodesics, how they interact with incoming particles or their stability under perturbations. So far, this type of analysis has only been performed for two-charge microstate geometries or three-charge geometries with atypical asymptotic charges and angular momentum\footnote{See \cite{Bena:2017upb} for a first approach to the study of such properties in superstrata microstate geometries, which have arbitrarily small angular momentum but are technically hard to describe and examine.}  \cite{Eperon:2017bwq, Eperon:2016cdd, Keir:2016azt, Marolf:2016nwu}.

As the procedure described is systematic, it would be very interesting to apply the tools developed in \cite{Heidmann:2017cxt} to perform macroscopic explorations of the parameter space. For instance, in \cite{Bena:2017fvm} this type of analysis has been successfully used to study generic four-center axisymmetric configurations, which can be constructed systematically, showing that those can only reproduce solutions with an angular momentum larger than $80\%$ of the cosmic censorship bound when they are smooth in five dimensions, while it is possible to find solutions with arbitrarily small angular momentum if the configuration contains a supertube (which are smooth only in six dimensions or more). Making use of the method that we propose here, we can access the full space of parameters of multicenter, not necessarily axisymmetric, solutions. Work along these lines is in progress \cite{inprogress}.

%%%%%%%%%%%%%%%%%%%%%%%%%%%%%%%%%%%%%%%%%%%%%%%%%%%%%%%%%%%%%%%%
\section*{Acknowledgments}

We are grateful to Iosif Bena, Pablo A. Cano, José J. Fernández-Melgarejo, Pierre Heidmann, \'Oscar Lasso, Carlos S. Shahbazi and Tom\'as Ort\'in for their very useful advice. This work has been supported in part by the MINECO/FEDER, UE grant
FPA2015-66793-P and by the Spanish Research Agency (Agencia Estatal de
Investigación) through the grant IFT Centro de Excelencia Severo Ochoa
SEV-2016-0597. The work of PFR was supported by the Severo Ochoa pre-doctoral grant SVP-2013-067903, the work of JA is supported by the Spanish Ministry of Economy, Industry and Competitiveness grant no. FIS2015-64654-P.

%%%%%%%%%%%%%%%%%%%%%%%5

%%%%%%%%%%%%%%%%%%%%5

\appendix

\section{Regular, horizonless solutions of SEYM theories}
\label{app:A}

In this appendix we fix our notation and summarize the construction of microstate geometries. We start with a very brief description of SEYM theories and continue in section \ref{app:solutions} with a summary of the results of \cite{Meessen:2015enl}, but in slightly different conventions. In appendix \ref{app:pelusos} we describe the construction of microstate geometries, adapting the results of \cite{Ramirez:2016tqc} to our current conventions, which have chosen to make contact with most of the literature on five-dimensional microstate geometries. Finally appendix \ref{app:charges} contains the expressions for the asymptotic charges in terms of the parameters of the solutions.

\subsection{Theory and conventions}

SEYM theories are $\mathcal{N}=1$, $d=5$ supergravities in which a non-Abelian subgroup, typically $SU(2)$, of the isometries of the scalar manifold has been gauged. For a thoughtful description of these theories we recommend the magnificent book \cite{Ortin:2015hya}. We set all fermions to zero and consider the bosonic part of the action,

\begin{equation}
\small
\label{eq:action}
\begin{array}{rcl}
S & = &  {\displaystyle\int} d^{5}x\sqrt{\vert g\vert}\
\biggl\{
R
+{\textstyle\frac{1}{2}}g_{xy}\mathfrak{D}_{\mu}\phi^{x}
\mathfrak{D}^{\mu}\phi^{y}
-{\textstyle\frac{1}{4}} a_{IJ} F^{I\, \mu\nu}F^{J}{}_{\mu\nu}
-\tfrac{1}{4}C_{IJK}
{\displaystyle\frac{\varepsilon^{\mu\nu\rho\sigma\lambda}}{\sqrt{\vert g \vert}}}
\left[
F^{I}{}_{\mu\nu}F^{J}{}_{\rho\sigma}A^{K}{}_{\lambda}
\right.
\\ \\ & & 
\left.
-\tfrac{1}{2}{g} f_{LM}{}^{I} F^{J}{}_{\mu\nu} 
A^{K}{}_{\rho} A^{L}{}_{\sigma} A^{M}{}_{\lambda}
+\tfrac{1}{10} g^2 f_{LM}{}^{I} f_{NP}{}^{J} 
A^{K}{}_{\mu} A^{L}{}_{\nu} A^{M}{}_{\rho} A^{N}{}_{\sigma} A^{P}{}_{\lambda}
\right]
\biggr\} \, ,
\end{array}
\end{equation}

\noindent
that describes the coupling of the metric, $n_v$ scalars labeled as $x,y=1,\dots,n_v$ and $(n_v+1)$ vector fields labeled with the indices $I,J, \ldots =0,\dots,n_v$. The full theory is completely determined by the election of the constant symmetric tensor $C_{IJK}$. We consider the $SU(2)$-gauged $ST[2,6]$ model, that contains $n_v=5$ vector multiplets. This model is characterized by a constant symmetric tensor with the following non-vanishing components

\begin{equation}
C_{0xy} = \frac{1}{6} \left( \begin{matrix}
    0 & 1 & 0 & 0 & 0 \\
    1 & 0 & 0 & 0 & 0 \\
    0 & 0 & -1 & 0 & 0 \\
    0 & 0 & 0 & -1 & 0 \\
    0 & 0 & 0 & 0 & -1
  \end{matrix}
  \right) \, .
\end{equation}

\noindent
The first three vectors, $A^0$, $A^1$ and $A^2$ are Abelian, while $A^3$, $A^4$ and $A^5$ correspond to a $SU(2)$ triplet. For convenience, we separate the range of values of the indices $I,J$ in two sectors: the Abelian sector $i,j=0,1,2$ and the non-Abelian sector $\alpha,\beta=3,4,5$. Therefore, if the latter sector is truncated we immediately recover the STU model of supergravity, with $C_{ijk}=\vert \varepsilon_{ijk} \vert /6$, the theory in which five-dimensional BPS microstate geometries are naturally described. 

It is convenient to introduce $(n_v+1)$ functions of the physical scalars $h^I(\phi^x)$, which are subjected to the following constraint

\begin{equation}
C_{IJK} h^I h^J h^K = 1 \, .
\end{equation}

\noindent
The functions $h^I$ can be interpreted as coordinates in a $(n_v+1)$-dimensional ambient space, so the above constraint defines a codimension 1 hypersurface parametrized by the scalars $\phi^x$ known as the scalar manifold. In the $ST[2,6]$ model, a convenient parametrization is

\begin{equation}
h^0 = e^{- \phi}e^{2k/3}  , \hspace{0.3cm}
h^1= \sqrt{2} e^{-4k/3}  , \hspace{0.3cm}
h^2= \sqrt{2} e^{-4k/3} \left( \vec{l}^2 + \frac{1}{2}e^\phi e^{2k} \right)  , \hspace{0.3cm}
h^{3,4,5} = -2e^{-4k/3} l^{3,4,5} ,
\end{equation}

\noindent
where the physical scalars coincide with the Heterotic dilaton $e^\phi$, the Kaluza-Klein scalar $e^k$ of the dimensional reduction from six to five dimensions and the non-Abelian scalars $l^\alpha$ appearing in the reduction of the vectors. 

We also define

\begin{equation}
h_I \equiv \frac{\partial}{\partial h^I} C_{JKL} h^J h^K h^L= 3 C_{IJK} h^J h^K \, , \qquad h_I = a_{IJ} h^J \, .
\end{equation}

\noindent
The matrix $a_{IJ}$ is the metric in the ambient space, and the $\sigma$-model metric $g_{xy}$ in the action is given by the pullback of $a_{IJ}$ on the hypersurface. They are both determined by the election of $C_{IJK}$ as

\begin{equation}
a_{IJ}=-6 C_{IJK} h^K + h_I h_J \, , \qquad g_{xy} = a_{IJ} \frac{ \partial h^I }{\partial \phi^x} \frac{\partial h^J}{\partial \phi^y} \, .
\end{equation}

\noindent
We only consider symmetric scalar manifolds, for which

\begin{equation}
\label{eq:symscalar}
C^{IJK} h_I h_J h_K=1 \, , \qquad
h^I=3 C^{IJK} h_J h_K \, , \qquad
\text{with} \, \, \, C^{IJK} \equiv C_{IJK} \, .
\end{equation}

The field strength and covariant derivatives are defined in the usual manner,

\begin{equation}
F^I\,_{\mu \nu} = 2\partial_{[\mu} A^I\,_{\nu]} + g f_{JK}\,^I A^J\,_\mu A^K\, _\nu \, , \qquad
\mathfrak{D}_\mu \phi^x=\partial_\mu \phi^x+g A^\alpha \,_\mu k_\alpha\,^x \, .
\end{equation}

\noindent
We consider the gauge group $SU(2)$ with structure constants $f_{IJ}\, ^K=\varepsilon_{IJ}\,^K$, with the understanding that they vanish whenever any of the indices takes values in the Abelian sector. The covariant derivatives of the functions of the scalars are

\begin{equation}
\mathfrak{D}_\mu h^I = \partial_\mu h^I+g f_{JK}\,^I A^J h^K \, , \qquad
\mathfrak{D}_\mu h_I = \partial_\mu h_I+g f_{IJ}\,^K A^J h_K \, .
\end{equation}

%%%%%%%%%%%%%%%%%%%%%%%%%%%%%%%%%%%%%%%%%%%%
\subsection{Timelike supersymmetric solutions with one isometry}
\label{app:solutions}

Supersymmetric solutions of this theory admit a Killing vector of non-negative norm. In adapted coordinates the metric and vectors are independent of the time coordinate, see \cite{Bellorin:2007yp}, and can be written as
\begin{eqnarray}
\label{eq:themetric}
ds^{2} 
&=& 
f^{\, 2}(dt+\omega)^{2}
-f^{\, -1}d\hat{s}^{2}\, , \\
& & \nonumber \\
\label{eq:vectorfields}
A^{I} 
& = &
h^{I} f (dt +\omega) +\hat{A}^{I}\, ,
\end{eqnarray}

\noindent
where $d\hat{s}^2$ is a hyperK\"ahler metric. The equations of motion are reduced to the following BPS system of differential equations on this four-dimensional space,

\begin{eqnarray}
\hat{F}^I &=& \star_4 \hat{F}^I \, , \\
& & \nonumber \\
\label{eq:BPS}
\hat{\mathfrak{D}}^2 Z_I &=& 3 \, C_{IJK} \star_4 \left( \hat{F}^J \wedge \hat{F}^K \right) \, , \\
& & \nonumber \\
d\omega + \star_4 d\omega &=&  Z_I \hat{F}^I \, ,
\end{eqnarray}

\noindent
where $\star_4$ is the Hodge dual in the hyperKähler space, $\hat{F}^I$ is the field strength of the vector $\hat{A}^I$ and $\hat{\mathfrak{D}}$ is the covariant derivative with connection $\hat{A}$. In these equations we have introduced the functions $Z_I \equiv h_I/f$, so the metric function $f$ is conveniently obtained as

\begin{equation}
f^{-3} = C^{IJK} Z_I Z_J Z_K \, ,
\end{equation}

\noindent
by virtue of equation \eqref{eq:symscalar}.

The system of BPS equations is non-linear due to the presence of non-Abelian fields, although the three equations could be solved independently in the order they have been presented. However, it is possible to further simplify the system under the assumption that the solution admits a spacelike isometry \cite{Meessen:2015enl}, in a way that reduces the problem to a set of equations in three dimensional Euclidean space. First, consider the following decompositions

\begin{eqnarray}
\label{eq:GHmetric}
d\hat{s}^{2}
&=&
H^{-1} (d\psi +\chi)^{2}
+H dx^s dx^s \, , \\
& & \nonumber \\
\label{eq:instantondec}
\hat{A}^{I}
& = &
-H^{-1}\Phi^{I} (d\psi+\chi)+\breve{A}^{I}\, , \\
& & \nonumber \\
\label{eq:hIf}
Z_I
&=& 
L_{I}+3C_{IJK}\Phi^{J}\Phi^{K}H^{-1}\, , \\
& & \nonumber \\
\label{eq:omegahat}
\omega 
& = & 
\omega_{5}(d\psi+\chi) +\breve{\omega}\, ,
\end{eqnarray}

\noindent
where $\psi$ is the coordinate adapted to the spatial isometry. When these expressions are substituted in the BPS system of equations, we obtain the following simplified system of differential equations and algebraic relations

\begin{eqnarray}
\label{eq:Hequation}
\star_{3}dH 
& = &
d\chi  \, ,
\\
& & \nonumber \\
\label{eq:PhiIequation} 
\star_{3}\breve{\mathfrak{D}} \Phi^{I}
& = &
\breve{F}^{I}\, ,
\\
& & \nonumber \\
\label{eq:LIequation}
\breve{\mathfrak{D}}^{2}L_{I} 
& = &
{g}^{2} f_{IJ}{}^{L}f_{KL}{}^{M}\Phi^{J}\Phi^{K}L_{M}\, ,  
\\
& & \nonumber \\
\label{eq:omegaequation}
\star_{3} d \breve{\omega}
&=&
H dM-MdH
+\frac{1}{2} ( \Phi^{I} \breve{\mathfrak{D}}L_{I}
-L_{I}\breve{\mathfrak{D}} \Phi^{I} )
\, , \\
& & \nonumber \\
\label{eq:omega5}
\omega_5
&=&
M+\frac{1}{2} {L_I \Phi^{I}}{H^{-1}}+ C_{IJK} {\Phi^{I} \Phi^{J} \Phi^{K}}{H^{-2}} \, ,
\end{eqnarray}

\noindent
where $\breve{F}^I$ is the field strength of the vector $\breve{A}^I$ and $\breve{\mathfrak{D}}$ is the covariant derivative with connection $\breve{A}$.

The Abelian functions $H$, $M$, $\Phi^i$ and $L_i$ are just harmonic functions in $\mathbb{E}^3$, and the $1$-forms $\chi$ and $\breve{A}^i$ are completely determined from those functions. In the non-Abelian sector, equations (\ref{eq:PhiIequation}) are non-linear and must be solved simultaneously for $\Phi^\alpha$ and $\breve{A}^\alpha$, which make their presence in (\ref{eq:LIequation}). The construction of non-Abelian microstate geometries requires finding a multicenter solution to these equations. The only known example of such solution is the multicolored dyon, found by one of us in \cite{Ramirez:2016tqc}, which we review in appendix~\ref{app:pelusos}. Last but not least, we have the differential equation (\ref{eq:omegaequation}), whose integrability condition will give rise to the bubble equations. 

Notice that these solutions are left invariant under the following transformations of the harmonic functions generated by the parameters $g^i$, whose sole effect is a gauge transformation of the Abelian vectors,

\begin{equation}
\begin{aligned}
&H' = H, \qquad \Phi^{i\, '} = \Phi^{i} \:+\: g^i H, \\
&L_{i} ' = L_{i} \:-\: 6 C_{ijk}\, g^j \Phi^k \:-\: 3 C_{ijk}\, g^j g^k H, \\
&M'=  M \:-\: \frac{1}{2} g^i L_i \:+\: \frac{3}{2} C_{ijk}\, g^i g^j \Phi^k \:+\: \frac{1}{2} C_{ijk} \,g^i g^j g^k H,
\label{eq:gaugetransformation}
\end{aligned}
\end{equation}

%%%%%%%%%%%%%%%%%%%%%%%%%%%%%%%%%%%%%%%%%%%%%%%%%%%%%%%%%%%%
\subsection{Microstate geometries in a nutshell}
\label{app:pelusos}

The previous section describes a procedure to find supersymmetric solutions of SEYM theories in terms of a set of three-dimensional \emph{seed functions}: $H, M, \Phi^I$ and $L_I$. As we already commented, those in the Abelian sector are just multicenter harmonic functions with poles in a collection of $n$ points located at $(x^1_a,x^2_a,x^3_a)$ called centers,

\begin{equation}
H=\sum_{a=1}^n \frac{q_a}{r_a} \, , \qquad
\Phi^i=\sum_{a=1}^n \frac{k^i_a}{r_a} \, , \qquad
L_i=l^i_0+\sum_{a=1}^n \frac{l^i_a}{r_a} \, , \qquad
M=m_0+\sum_{a=1}^n \frac{m_a}{r_a} \, ,
\end{equation}

\noindent
with $r_a=\vert \vec{x} - \vec{x}_a \vert$. Notice that these functions solve the equations \eqref{eq:Hequation}-\eqref{eq:LIequation} in the Abelian sector everywhere except at the locations of the poles. This is the reason why the bubble equations are needed.

In the non-Abelian sector, the Bogomoln'yi equations \eqref{eq:PhiIequation} can be readily solved by making use of the following ansatz

\begin{equation}
\Phi^\alpha= -\frac{1}{gP} \frac{\partial P}{\partial x^s} \delta^\alpha_s \, , \qquad
\breve{A}^\alpha\,_\mu = - \frac{1}{gP} \frac{\partial P}{\partial x^s} \varepsilon^\alpha\,_{\mu s} \, .
\end{equation}

\noindent
Obtaining the condition for the function 

\begin{equation}
\frac{1}{P} \nabla^2 P = 0 \, ,
\end{equation}

\noindent
which is solved again by a harmonic function $P$, even at the locations of the poles. Equations \eqref{eq:LIequation} for the non-Abelian sector can also be solved using the ansatz

\begin{equation}
L_\alpha= -\frac{1}{gP} \frac{\partial Q}{\partial x^s} \delta_\alpha^s  \, , 
\end{equation}

\noindent
which yields

\begin{equation}
\frac{\partial}{\partial x^s} \left(\frac{1}{P^2} \nabla^2 Q \right)= 0 \, .
\end{equation}

\noindent
This condition is solved everywhere if $Q$ is a harmonic function with the poles at the same locations than $P$. Therefore, the complete non-Abelian multicolored dyon is specified by two harmonic functions\footnote{We assume that the constant term of the function $P$ is non-vanishing, in which case it can always be taken to be $1$. From the Bogomol'nyi equation perspective, truncating this constant is equivalent to adding a unit charge monopole at infinity. We leave the study of this possibility for future works.}

\begin{equation}
P=1+\sum_{a=1}^n \frac{\lambda_a}{r_a} \, , \qquad 
Q=\sum_{a=1}^n \frac{\sigma_a \lambda_a}{r_a} \, , \qquad
\text{with} \, \, \, \lambda_a >0 \, .
\end{equation}

In order to avoid the presence of event horizons or singularities at the centers, it is necessary to fix the value of some of the parameters,

\begin{equation}
\label{eq:fixparameters}
l^0_a=-\frac{1}{q_a}\left(k^1_ak^2_a - \frac{1}{2 g^{2}} \right) \, , \qquad
l^{1,2}_a=- \frac{k^0_a k^{2,1}_a}{q_a} \, , \qquad
\sigma_a= \frac{k^0_a}{q_a} \, , \qquad
m_a= \frac{k^0_a}{2 q_a^2}\left(k^1_a k^2_a-\frac{1}{ 2 g^{2}}\right) \, .
\end{equation}

\noindent
On its side, asymptotic flatness requires

\begin{equation}
\label{eq:fixasymptotics}
l^0_0 l^1_0 l^2_0 = 1 \, , \qquad
m_0 = - \frac{1}{2} \sum_{i,a} l^i_0 k^i_a \, .
\end{equation}

The integrability condition of equation \eqref{eq:omegaequation} gives the set of constraints known as bubble equations

\begin{equation}
\sum_{b\neq a} \frac{q_aq_b}{r_{ab}}\Pi^0_{ab}\left(\Pi^1_{ab} \Pi^{2}_{ab} - \frac{1}{2g^{2}}\mathbb T_{ab}\right)=\sum_{b,i} q_a q_b l_0^i  \Pi^i_{ab} \, .
\end{equation}

\noindent
where

\begin{equation}
\Pi^i_{ab} \equiv \frac{1}{4\pi} \int_{\Delta_{ab}} F^i = \left( \frac{k^i_b}{q_b}-\frac{k^i_a}{q_a} \right) \, , \qquad
\mathbb{T}_{ab} \equiv \left( \frac{1}{q_a^2}+\frac{1}{q_b^2} \right) \, .
\end{equation}

\noindent
The term $\mathbb{T}_{ab}$ appears due to the presence of non-Abelian fields, that alter the value of the parameters $l^0_a$ when compared with purely Abelian configurations. The $i$-fluxes threading the non-contractible 2-cycles $\Delta_{ab}$ defined by any path connecting two centers $\vec{x}_a$ and $\vec{x}_b$ behave effectively as sources of electric charge and mass. When all the bubble equations are satisfied, the solutions are regular at the centers and do not present Dirac-Misner string singularities, which otherwise could only be removed by compactifying the time direction.  

The last restriction for the construction of physically sensible microstate geometries comes from demanding that the solution does not contain closed timelike curves (CTCs). The metric can be rewritten in the following manner

\begin{equation}
\label{eq:metricexp}
ds^2= f^2 dt^2 +2f^2 dt \omega-\frac{\mathcal{I}_4}{f^{-2} H^2} \left( d\psi + \chi - \frac{\omega_5 H^2}{\mathcal{I}_4} \breve{\omega} \right)^2 - f^{-1} H \left( d\vec{x} \cdot d\vec{x}-\frac{\breve{\omega}^2}{\mathcal{I}_4} \right) \, ,
\end{equation}

\noindent
where $\mathcal{I}_4$ is the \emph{quartic invariant}, defined as

\begin{equation}
\mathcal{I}_4 \equiv f^{-3} H - \omega_5^2 H^2 \, .
\end{equation}

\noindent
Therefore, a general restriction that must be satisfied in order to avoid CTCs is the positivity of the quartic invariant

\begin{equation}
\label{eq:CTC1}
\mathcal{I}_4 \geq 0 \, .
\end{equation}

\noindent
When studying its positivity numerically, it is sometimes useful to employ the expression for the quartic invariant in terms of the seed functions directly
\begin{equation}
\label{eq:CTC}
\begin{array}{rcl}
\mathcal{I} &=& -M^2 H^2- \frac{1}{4} \left( \Phi^I L_ I \right)^2 - 2 M C_{IJK} \Phi^I \Phi^J \Phi^K - M H L_ I \Phi^I \\ \\
& & + H C^{IJK} L_I L_J L_K + 9 C^{IJK} C_{KLM} L_I L_J \Phi^L \Phi^M \geq 0 \, .
\end{array}
\end{equation}

%%%%%%%%%%%%%%%%%%%%%%%%%%%%%%%%%%%%%%%%%%%%%%%%%%%
\subsection{Asymptotic charges}
\label{app:charges}

The electric asymptotic charge of each Abelian vector can be readily obtained from the asymptotic expansion of the associated warp factor, see \cite{Bena:2007kg},

\begin{equation}
Z_{i,\infty} = l^i_0 + \frac{\mathcal {Q}_i}{r} + \mathcal{O}(r^{-2}) \, .
\end{equation}

\noindent
This can be easily seen from the fact that, asymptotically, $A^i\,_{t,\infty} \sim Z^{-1}_i$. The electric charges are

\begin{eqnarray}
\label{eq:charges}
\mathcal Q_0 &=& -\sum_{a,b,c} q_a q_b q_c \Pi^1_{ab} \Pi^2_{ac}+ \frac{1}{2g^2}\sum_a \frac{1}{q_a} \, , \\
\mathcal Q_1 &=& -\sum_{a,b,c} q_a q_b q_c \Pi^0_{ab} \Pi^2_{ac} \, ,\\
\mathcal Q_2 &=& -\sum_{a,b,c} q_a q_b q_c \Pi^0_{ab} \Pi^1_{ac} \, .
\end{eqnarray} 

\noindent
In a similar manner, the two angular momenta can be read from the term in $d\psi$ at the asymptotic expansion of $\omega$ \cite{Bena:2007kg}, whose contribution comes entirely from the function $\omega_5$, obtaining

\begin{eqnarray}
J_R &=& - \frac{1}{2} \sum_{a,b,c,d} q_a q_b q_c q_d \Pi^0_{ab} \Pi^1_{ac} \Pi^2_{ad} + \frac{1}{4g^2} \sum_{a,b} \frac{q_b \Pi^0_{ab}}{q_a} \, , \\
\vec{J}_L &=& - \frac{1}{4} \sum_{\substack{a,b \\ a\neq b}} q_a q_b\Pi^0_{ab}\left(\Pi^1_{ab} \Pi^{2}_{ab} - \frac{1}{2g^{2}}\mathbb T_{ab}\right) \frac{\vec{x}_a-\vec{x}_b}{\vert \vec{x}_a-\vec{x}_b \vert} \, .
\end{eqnarray}

\noindent
such that

\begin{equation}
\omega_{5,\infty} = \frac{1}{r}\left({J}_R + {J}_L \cos \theta_L \right)  +  \mathcal{O}(r^{-2}) \, ,
\end{equation}

\noindent
where $\theta_L$ is the angle measured with respect to $\vec{J}_L$, and $J_L$ is the norm of this vector.

The ADM mass is just

\begin{equation}
\mathcal M=\frac{\pi}{G_N^{(5)}}\left(\frac{\mathcal Q_0}{l_0^0}+\frac{\mathcal Q_1}{l_0^1}+\frac{\mathcal Q_2}{l_0^2}\right) \, .
\end{equation}

If all the centers were placed at the same location, the solution would describe a black hole with the same asymptotic charges (with $J_L=0$) and an event horizon whose area would be given by

\begin{equation}
A_H=2\pi^2\sqrt{\mathcal Q_0 \mathcal Q_1 \mathcal Q_2 -J_R^2} \, .
\end{equation}

\noindent
It is convenient to define the \emph{entropy parameter} $\mathcal{H}$ of a microstate geometry as

\begin{equation}
\mathcal{H} \equiv 1- \frac{J^2_R}{\mathcal Q_0 \mathcal Q_1 \mathcal Q_2} \, ,
\end{equation}

\noindent
whose value indicates how far from maximal rotation the represented black hole is.

\renewcommand{\leftmark}{\MakeUppercase{Bibliography}}
\phantomsection
\bibliographystyle{JHEP}
\bibliography{references}
\label{biblio}

\end{document}